\documentclass[5p]{elsarticle}
\usepackage{amssymb}
\usepackage{amsmath}
\usepackage{graphicx}
\usepackage{natbib}
\usepackage{txfonts}

\bibpunct{(}{)}{;}{a}{,}{,}
\journal{New Astronomy}
\newcommand{\myr}{\,$M_{\odot}\,{\rm yr}^{-1}$}
\newcommand{\lo}{\,$L_{\odot}$}
\def\astrobj#1{#1}
\begin{document}
\begin{frontmatter}

\title{Notices to investigation of symbiotic binaries\\
       V. Physical parameters derived from $UBV$ magnitudes}

\author{Z.~Carikov\'{a}}
\ead{zcarikova@ta3.sk}

\author{A.~Skopal}
\ead{skopal@ta3.sk}

\address{Astronomical Institute, Slovak Academy of Sciences,
        059~60 Tatransk\'{a} Lomnica, Slovakia \\[2mm]
        {\rm Received -; accepted -}}

\begin{abstract}
In the optical, the spectrum of symbiotic binaries consists
of contributions from the cool giant, symbiotic nebula and the
hot star. Strong emission lines are superposed on the continuum.
In this paper we introduce a simple method to extract individual
components of radiation from photometric $UBV$ magnitudes.
We applied the method to classical symbiotic stars \astrobj{AX Per},
\astrobj{AG Dra}, \astrobj{AG Peg} and \astrobj{Z And},
the symbiotic novae \astrobj{RR Tel} and \astrobj{V1016 Cyg} and
the classical nova \astrobj{V1974 Cyg} during its nebular phase.
We estimated the electron temperature and emission measure of
the nebula in these systems and the $V$ magnitude of the
giant in the symbiotic objects.
Our results are in a good agreement with those obtained
independently by a precious modelling the UV--IR SED.
\end{abstract}
\end{frontmatter}
\section{Introduction}
Symbiotic stars are long-period ($P_{orb}\sim$ 1 -- 3 years or more)
interacting binary systems, which comprise a late-type giant
(usually a red giant of the spectral type M) and a hot compact
star, most probably a white dwarf. If the giant of the spectral
type G or K is present, we call them 'yellow symbiotics'.
During quiescent phases, the white dwarf accretes a fraction of
the wind from the giant. Typical mass loss rate from the giant
in the symbiotic binaries is of a few times $10^{-7}$\myr.
The accretion process leads to heating up the white dwarf's
surface to a very high temperature of $T_{h}\sim 10^{5}$\,K,
and increases its luminosity to $L_{h} \sim 10^{2}-10^{4}$\lo.
Such the hot and luminous WD is capable of ionizing
neutral wind particles of both the stars giving rise
to a strong nebular radiation.
During quiescent phases, the symbiotic nebula represents
mostly the ionized part of the stellar wind from the giant.
During active phases, the mass loss rate from the hot
component increases, and can temporary exceed that from
the giant. As a result, properties of the symbiotic nebula
significantly change during the active phases.

Accordingly, the observed flux, $F_{\lambda}$ (corrected for
the interstellar extinction) is given by the superposition of
three basic components of radiation -- from the nebula,
$F_{\lambda}^{nebula}$, the cool giant, $F_{\lambda}^{giant}$,
and the hot stellar source, $F_{\lambda}^{hot}$, 
i.e.
\begin{equation}
 F_{\lambda} = F_{\lambda}^{nebula} + F_{\lambda}^{giant} + 
               F_{\lambda}^{hot}.
\label{eqn:totalflux}
\end{equation}
Contributions of the individual components of the radiation
depend on the wavelength. The hot stellar source dominates
the far-$UV$ spectrum, while the nebula is usually dominant
in the near-$UV$/$U$ domain, and the cool giant (especially
the red giant) dominates the near-$IR$ spectral region. In the
case of yellow symbiotics the contribution of the giant
to the Johnson's $V$ filter is very strong relatively to that
from the nebula. However, contributions from different
sources in a symbiotic system depend also on the level of
the activity.

During active phases of symbiotic binaries with a high orbital
inclination, we observe narrow minima in their light
curves. They can be interpreted as eclipses of the hot object
by the cool giant. This implies that during the activity
the hot stellar source produces a dominant amount of its
radiation within the optical region. It is believed that an
optically thick disk is formed around the hot star at the orbital
plane during active phases \citep[see][]{sk05}.
Its relatively cool rim mimics a warm pseudophotosphere, while
the circumstellar material above/below the disk can easily be
ionized by its hot central part. Therefore the observed depths
of the minima are partially filled in by the nebular radiation.
Thus during the eclipses, the observed light consists
of the contributions from the giant and the nebula.
During quiescent phases the optical region is dominated by
the extensive nebula, which thus prevents from observing
eclipses. Corresponding colour indices are very different
from those observed during active phases.

The aim of this paper is to propose a method of disentangling
photometric $UBV$ magnitudes of symbiotic stars into their
components from contributing sources as given by
Eq. (\ref {eqn:totalflux}).
This means to determine physical parameters of their radiation.
In particular, the electron temperature, $T_{\rm e}$, and emission
measure, $EM$, of the symbiotic nebula, and the brightness of
the cool giant in the $V$ passband ($V^{giant}$).
We describe our method in the following section. First, we
formulated some simplifying assumptions (Sect.~2.1), and then
we derived a system of equations, which relates theoretical
properties of the radiative components to their observational
characteristics (Sect.~2.4). In Sect.~3 we applied this approach
to selected objects and compared the results with those achieved
independently by another methods. Discussion and summary of
the results are found in Sects.~4 and 5, respectively.
\section{The method}

\subsection{Assumptions}

In our approach we assume that the contribution from the hot
stellar source can be neglected within the optical. We thus
consider contributions only from the nebula and the cool giant,
which simplifies Eq. (\ref {eqn:totalflux}) to
  $F_{\lambda} = F_{\lambda}^{nebula} + F_{\lambda}^{giant}$.
This assumption can be applied for systems during quiescent
phases, while during active phases, it is valid only under
specific conditions. In the following two points we summarize
reasons and conditions, under which this assumption can
be applied.

(i)
During quiescent phases the hot star contribution can be
neglected in the optical, because of its very high temperature,
which moves the maximum of its radiation to the extreme ultraviolet
or supersoft X-ray domain, while the optical is contributed
only with a very faint Rayleigh-Jeans tail in a blackbody
approximation \citep[e.g.][]{kw84,m+91,sk05}. We investigated
this case for three classical symbiotic stars \astrobj{AG Dra},
\astrobj{AG Peg} and \astrobj{Z And} during their quiescent phases.

(ii)
During active phases, a large optically thick disk is created
around the accretor at the orbital plane. Its signatures are
indicated best for the systems with a high orbital inclination
\citep[see Fig.~27 of][]{sk05}. For the eclipsing symbiotics the
hot stellar source is represented by a relatively warm
($T_{\rm h} \sim 22\,000$\,K) disk's rim, which light contribution
can be significant within the optical region. Therefore, the
radiation from the hot active object can be neglected only during
the eclipses. As example here, we analysed the light from
the 1994 eclipse of \astrobj{AX Per} during its 1989-95 active phase.
In the case of non-eclipsing systems, the optical light from
the hot object can be neglected at any orbital phase, because
we are viewing the active hot star more from its pole, and thus
its radiation is not cooled by the disk material on the line
of sight. We demonstrate this case on example of \astrobj{AG Dra}. 

Finally, we note that our method is applicable to any object,
whose continuous spectrum consists of two or one (nebular)
component of radiation. We demonstrate this case on examples
of symbiotic novae \astrobj{RR Tel}, \astrobj{V1016 Cyg} and
a classical nova \astrobj{V1974 Cyg} during its nebular phase.

\subsection{The true continuum from $UBV$ magnitudes}

Photometric measurements i.e. observed $UBV$ magnitudes used in this
work are summarized in Table~\ref {table:inputubv}.

First, we corrected the observed magnitudes for the interstellar
extinction using the extinction curve of \citet[][]{ccm89} with
appropriate $E_{B-V}$ colour excess (Table~\ref{table:exdis}).
Second, to determine the true continuum using the multicolour
photometry, we corrected the dereddened magnitudes for the influence
of emission lines. For the purpose of this paper we used corrections
presented by \citet[][]{lines}, if not specified otherwise.
The corrections are summarized in Table~\ref {table:corr}.

Finally, according to the Pogson's equation, we converted the
magnitudes of the true continuum, $m_{\lambda}$, to fluxes by
\begin{equation}
 F_{\lambda} = 10^{-0.4(m_{\lambda}+q_{\lambda})},
\end{equation}
where the constant $q_{\lambda}$ defines the magnitude zero.
For the standard Johnson $UBV$ photometric system and the fluxes
in units of $\rm \,erg\,cm^{-2}\,s^{-1}\,\AA^{-1}$, $q_{\rm U} = 20.9$\,mag, $q_{\rm B} = 20.36$\,mag
and $q_{\rm V} = 21.02$\,mag \citep[e.g.][]{h+k82}.
%
\begin{table}
\caption[]{Colour excesses, distances and spectral types of the giants.}
\begin{center}
{\footnotesize
\begin{tabular}{ccccc}
\hline
\hline
Object      & $E_{B-V}$ & $d$   & Spectral & ref.\\
            & [mag]     & [kpc] & type &    \\
\hline
AX~Per      & 0.27 & 1.73 & M4.5 & 1,7\\
AG~Dra      & 0.08 & 1.1  & K2(K4) & 1,7\\
AG~Peg      & 0.10 & 0.80 & M3 & 1,7\\
 Z~And      & 0.30 & 1.5  & M4.5 & 1,7\\
RR Tel      & 0.10 & 2.5  & M6 & 2,3,7\\
V1016 Cyg   & 0.28 & 2.93 & M7 & 4,5,7\\
V1974 Cyg   & 0.32 & 1.77 & - & 6\\
\hline
\end{tabular}
}
\end{center}
{\scriptsize
References:\\
            1 - \citet[][]{sk05} and references therein\\
            2 - \citet[][]{penston83}\\
            3 - \citet[][]{whitelock87}\\
            4 - \citet[][]{nussschild81}\\
            5 - \citet[][]{parimucha03}\\
            6 - \citet[][]{chochol93}\\
            7 - \citet[][]{m+s99}\\
}						
\label{table:exdis}
\end{table}
\begin{table}[h]
\caption[]{Input parameters for investigated objects - Photometric measurements.}
\begin{center}
{\footnotesize
\begin{tabular}{ccccc}
\hline
\hline
Object & Julian date & U & B & V\\
       & $JD-2\,4...$ & [mag] & [mag] & [mag]\\
\hline
AX Per   & 49\,571 & 12.689 & 12.873 & 11.583\\
         & 49\,575 & 12.712 & 12.895 & 11.641\\
         & 49\,593 & 12.744 & 12.946 & 11.790\\
         & 49\,601 & 12.733 & 12.936 & 11.681\\
         & 49\,608 & 12.742 & 12.999 & 11.722\\
         & 49\,621 & 12.715 & 12.870 & 11.660\\
\hline
AG Dra   & 52\,765 & 10.957 & 11.029 & 9.735\\
         & 53\,178 & 10.952 & 11.044 & 9.720\\
         & 52\,919 &  9.189 & 10.022 & 9.181\\
\hline
AG Peg   & 49\,974 & 9.16 & 9.76 & 8.56\\
         & 49\,988 & 9.13 & 9.73 & 8.55\\
         & 49\,993 & 9.09 & 9.71 & 8.55\\
         & 49\,996 & 9.10 & 9.76 & 8.58\\
         & 50\,012 & 8.97 & 9.75 & 8.56\\
         & 50\,024 & 9.08 & 9.69 & 8.56\\
\hline
Z And    & 45\,172 & 11.41 & 11.72 & 10.50\\
         & 45\,195 & 11.28 & 11.74 & 10.64\\
         & 45\,196 & 11.27 & 11.77 & 10.64\\
         & 45\,204 & 11.24 & 11.77 & 10.67\\
\hline
RR Tel   & 48\,066 & 9.90 & 11.26 & 10.84\\
         & 48\,067 & 9.83 & 11.24 & 10.84\\
\hline
V1016 Cyg& 45\,975 & 10.26 & 11.41 & 11.15\\
\hline
V1974 Cyg& 48\,883 & 8.243 & 9.359 & 9.426\\
\hline
\end{tabular}
}
\end{center}
\label{table:inputubv}
\end{table}
\begin{table}[h]
\caption[]{Input parameters - Corrections for the emission lines.}
\begin{center}
{\footnotesize
\begin{tabular}{ccccc}
\hline
\hline
Object & Julian date & $\Delta U$ & $\Delta B$ & $\Delta V$\\
       & $JD-2\,4...$&   [mag]    &   [mag]    &   [mag]\\
\hline
AX Per   & 49\,600 & -0.77 & -0.62 & -0.19\\
\hline
AG Dra   & 52\,765 & -0.033& -0.10 & -0.01\\
         & 53\,172 & -0.030& -0.08 & -0.01\\
         & 52\,919 & -0.081& -0.12 & -0.02\\
\hline
AG Peg   & 49\,305 & -0.19 & -0.41 & -0.12\\
\hline
Z And    & 45\,293 & -0.30 & -0.48 & -0.12\\
\hline
RR Tel   & 51\,836 & -1.71 & -1.68 & -0.87\\
\hline
V1016 Cyg& 46\,045 & -1.23 & -1.67 & -1.19\\
\hline
V1974 Cyg& 48\,883 & -1.57 & -1.73 & -0.66\\
\hline
\end{tabular}
}
\end{center}
\label{table:corr}
\end{table}
%
\subsection{Contributions from the nebula and the giant}

Now, let us have a look on basic relations for the nebular continuum
and that from the giant. The nebular flux in
Eq. (\ref {eqn:totalflux}) can be approximated by
\begin{equation}
F_{\lambda}^{nebula} = k_{\rm n} \times \varepsilon_{\lambda}(T_{\rm e}),
\label{eqn:neb}
\end{equation}
where $k_{\rm n}$ [cm$^{-5}$] is the scaling factor, which determines
amount of the nebula and $\varepsilon_{\lambda}(T_{\rm e})$ is
the volume emission coefficient [${\rm erg\,cm^3\,s^{-1}\,\AA^{-1}}$],
which depends on the electron temperature of the nebula $T_{\rm e}$,
and is a function of the wavelength \citep[e.g.][]{b+m70}.
For the sake of simplicity, we calculated the volume emission
coefficient for the hydrogen plasma only, including contributions
from recombination and bremsstrahlung. In addition,
Eq. (\ref {eqn:neb})
requires $T_{\rm e}$, and thus the emission coefficient
$\varepsilon_{\lambda}(T_{\rm e})$, to be constant throughout
the nebula.
The total emission produced by the optically thin nebula is
\begin{equation}
4\pi d^2 F_{\lambda}^{nebula}\, =\,
\varepsilon_{\lambda}\int_V n_{\rm e}n_{+}{\rm d}V =
\varepsilon_{\lambda}\textsl{EM},
\end{equation}
where \textsl{EM} [cm$^{-3}$] is
the so-called emission measure. It is determined by the volume
of the nebula and concentrations of electrons
and ions (protons), $n_{\rm e}$ and $n_{+}$, respectively.
So, with the aid of Eq. (\ref {eqn:neb}), the emission measure
can be expressed as
\begin{equation}
 EM = 4 \pi d^{2}k_{n} = 4 \pi d^{2}\frac
                         {F_{\lambda}^{nebula}}
                         {\varepsilon_{\lambda}(T_{\rm e})},
\label{eqn:em} 
\end{equation} 
where $d$ is the distance to the object.

Due to the asymmetry of the $U$ filter with respect to the
wavelength of the Balmer discontinuity, $\lambda_{\rm Balmer}$,
we calculated the emission coefficient $\varepsilon_{\rm U}$ as
the weighted average of its values from both sides of the
Balmer jump \citep[see][]{N4}. For the response function
of the $U$ filter, as published by \citet[][]{mat+san63},
we get
\begin{equation}
 \varepsilon_{\rm U} \doteq 0.6\, \varepsilon_{\rm U^{-}} + 
                       0.4\, \varepsilon_{\rm U^{+}},
\end{equation}
where $\varepsilon_{\rm U^{-}}$ and $\varepsilon_{\rm U^{+}}$ are
emission coefficients at the short and long wavelength side of
the $\lambda_{\rm Balmer}$, respectively.

The stellar radiation from cool giants of different spectral
types can be characterized by different colour indices.
For the purpose of this work we used colour indices from
\citet [][]{johnson66}. To simplify relations introduced
in the following section, we denote the colour indices
of the giant's spectrum, $U-B$ and $B-V$, by $UB$ and $BV$,
respectively.

\subsection{Parameters of the nebula and giant}

\begin{figure}[h!]
\begin{center}
\resizebox{\hsize}{!}{\includegraphics[angle=270]{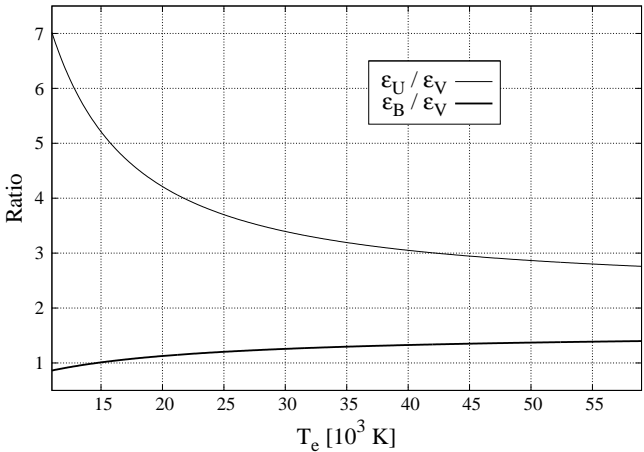}}
\caption[]{Ratios of the emission coefficients we used to solve 
           Eq. (\ref{eqn:te}).}
\label{fig:ee}
\end{center}
\end{figure}

Here we propose a method to estimate parameters of the nebular
radiation ($T_{\rm e}$, $EM$) and that of the giant (e.g. $V^{giant}$),
which dominates the optical part of the spectrum of most
symbiotic stars.
Based on the assumptions formulated in Sect.~2.1. and relations
for the nebular emission (Sect.~2.3.), we can write following set
of equations for the considered components of radiation.
First, the total flux in the continuum is given by the superposition
of fluxes from the nebula and the cool giant, i.e.
\begin{equation}
 F_{\rm U}^{nebula} + F_{\rm U}^{giant} = 10^{-0.4 (U^{cont} +q_{\rm U})},
\label{eqn:b}
\end{equation}
\begin{equation}
 F_{\rm B}^{nebula} + F_{\rm B}^{giant} = 10^{-0.4 (B^{cont} +q_{\rm B})},
\end{equation}
\begin{equation}
 F_{\rm V}^{nebula} + F_{\rm V}^{giant} = 10^{-0.4 (V^{cont} +q_{\rm V})},
\end{equation}
where $U^{cont}$, $B^{cont}$ and $V^{cont}$ are magnitudes of
the true continuum (Sect.~2.2.).
Second, according to Eq. (\ref {eqn:neb}), the nebular flux in
the spectrum is a function of the scaling factor $k_{\rm n}$
and the electron temperature $T_{\rm e}$, i.e.
\begin{equation}
 F_{\rm U}^{nebula} = k_{\rm n}~\varepsilon_{\rm U}(T_{\rm e}),
\end{equation}
\begin{equation}
 F_{\rm B}^{nebula} = k_{\rm n}~\varepsilon_{\rm B}(T_{\rm e}),
\end{equation}
\begin{equation}
 F_{\rm V}^{nebula} = k_{\rm n}~\varepsilon_{\rm V}(T_{\rm e}).
\end{equation}
Third, the ratio of the fluxes from the giant in different filters
can be expressed with the aid of the Pogson's equation as
\begin{equation}
 \frac{F_{\rm U}^{giant}}{F_{\rm B}^{giant}} = 
      10^{-0.4(U^{giant}-B^{giant}+q_{\rm U}-q_{\rm B})}=10^{-0.4(UB+q_{\rm U}-q_{\rm B})},
\end{equation}
\begin{equation}
 \frac{F_{\rm B}^{giant}}{F_{\rm V}^{giant}} = 
      10^{-0.4(B^{giant}-V^{giant}+q_{\rm B}-q_{\rm V})}=10^{-0.4(BV+q_{\rm B}-q_{\rm V})}.
\label{eqn:e}
\end{equation}
Thus we have a system of eight equations
(Eqs. (\ref {eqn:b}) to (\ref {eqn:e})) for eight variables,
\begin{equation}
\nonumber F_{\rm U}^{nebula},~F_{\rm B}^{nebula},~F_{\rm V}^{nebula},~F_{\rm U}^{giant},
         ~F_{\rm B}^{giant},~F_{\rm V}^{giant},~k_{\rm n}~{\rm and}~T_{\rm e}, 
\end{equation}
which can be solved for the known spectral type of the cool giant
(i.e. the $U-B$ and $B-V$ indices) and the measured $UBV$ magnitudes.
The aim is to determine the three fundamental parameters,
the electron temperature $T_{\rm e}$, the scaling factor $k_{\rm n}$
and the giant's magnitude, $V^{giant}$. The first two
parameters determine the nebular emission, while the third one
settles the radiation from the giant characterized with indices
$U-B$ and $B-V$, and thus defines the second term on the right side
of Eq. (\ref {eqn:totalflux}).
From the abovementioned system of equations 
(Eqs.~(\ref {eqn:b}) to (\ref {eqn:e})) we can derive a relation 
for determining the electron temperature in a form (see Appendix A) 
\begin{equation}
\begin{split}
 &\frac {\varepsilon_{\rm B}(T_{\rm e})}{\varepsilon_{\rm V}(T_{\rm e})}
     \left[10^{-0.4(V^{cont}+UB-q_{\rm B})}-10^{-0.4(U^{cont}-BV-q_{\rm B})}\right]+\\
 &+\frac {\varepsilon_{\rm U}(T_{\rm e})}{\varepsilon_{\rm V}(T_{\rm e})}
     \left[10^{-0.4(B^{cont}-BV-q_{\rm U})}-10^{-0.4(V^{cont}-q_{\rm U})}\right]+\\
 &+\left[10^{-0.4(U^{cont}-q_{\rm V})}-10^{-0.4(B^{cont}+UB-q_{\rm V})}\right]=0.
\end{split}
\label{eqn:te}
\end{equation}
Solving this equation for $T_{\rm e}$, allows us to determine
easily other parameters. For example, the scaling factor $k_{\rm n}$ as
\begin{equation}
 k_{\rm n}=\frac {F_{\rm V}^{nebula}}{\varepsilon_{\rm V}(T_{\rm e})}
 =\frac{10^{-0.4(BV+V^{cont}+q_{\rm B})}-10^{-0.4(B^{cont}+q_{\rm B})}}
 {10^{-0.4(BV+q_{\rm B}-q_{\rm V})}\varepsilon_{\rm V}(T_{\rm e})-\varepsilon_{\rm B}(T_{\rm e})}
\label{eqn:kn}
\end{equation}
and the giant's magnitude in the V passband $V^{giant}$ as
\begin{equation}
V^{giant}= -2.5 \log \left[10^{-0.4(V^{cont}+q_{\rm V})}-F_{\rm V}^{nebula} \right] -q_{\rm V},
\label{eqn:vg}
\end{equation}
where $F_{\rm V}^{nebula}$ is given by
\begin{equation}
 F_{\rm V}^{nebula}=\frac{10^{-0.4(BV+V^{cont}+q_{\rm B})}-10^{-0.4(B^{cont}+q_{\rm B})}}
 {10^{-0.4(BV+q_{\rm B}-q_{\rm V})}-\frac{\varepsilon_{\rm B}(T_{\rm e})}{\varepsilon_{\rm V}(T_{\rm e})}}.
\end{equation}
For more details see Appendix A.
The ratios of the emission coefficients 
$\varepsilon_{\rm U}/\varepsilon_{\rm V}$
and 
$\varepsilon_{\rm B}/\varepsilon_{\rm V}$ 
are plotted in Fig.~\ref {fig:ee}.

\section{Application to selected symbiotic stars}

Our resulting parameters, the electron temperature $T_{\rm e}$
(from Eq. (\ref{eqn:te})), scaling factor
$k_{\rm n}$ (Eqs. (\ref{eqn:neb}) and (\ref{eqn:kn})),
emission measure $EM$ (see Eq. (\ref{eqn:em})) and the $V$
magnitude of the giant $V^{giant}$ (Eq. (\ref{eqn:vg}))
are summarized in Table \ref{table:results}. Values of $EM$
were scaled for distances summarized in Table \ref{table:exdis}.
Disentangled $UBV$ magnitudes for the investigated objects
are plotted in Fig. 2.

Assuming that the uncertainties of the observed $U$, $B$ and $V$
magnitudes for all investigated systems are $\pm 0.05$\,mag,
$\pm 0.03$\,mag and $\pm 0.03$\,mag, respectively \citep[e.g.][]{sk+07},
we derived intervals of possible values of the searched
parameters, which can be found in the fourth, sixth, eighth
and tenth column in Table \ref{table:results}.
Uncertainties in the corrections for emission lines are
not included.
If more photometric measurements were available, we used mean
values of parameters, corrensponding to the individual $UBV$
magnitudes. For these cases, we adopted uncertainties
as the interception of parameter ranges given by individual
measurements
(except for the brightness of the cool giant in \astrobj{Z And}
and \astrobj{AX Per}
where we took the mean values of upper and lower limit of intervals
because those interceptions were empty).
Investigated objects are discussed in detail in the following
subsections.
\subsection{Classical symbiotic stars during quiescent phase}

\subsubsection{AG~Dra}

\astrobj{AG Dra} is a yellow symbiotic star, because its cool component
was classified as a normal K2 giant
\citep[e.g.][]{m+s99}.
To analyse a quiescent phase of \astrobj{AG Dra} by our method,
we used $U$, $B$, $V$ measurements made
on 2003/05/05 \citep[][]{photosymb11}
and 2004/06/21 \citep[][]{sk+07}. We selected these
dates, because of having simultaneous spectroscopic
observations, which allowed us to determine corrections for
emission lines on the $UBV$ magnitudes
\citep[Fig.~6 of][ and Viotti, private communication]{gon+08}.
The colour indices for a K2 giant,
$B-V = 1.16$ and $U-B = 1.16$ \citep[][]{johnson66}
are, however, in conflict with the observed
indices of true continuum,
$(B-V)_{\rm obs} \doteq 1.3$ and $(U-B)_{\rm obs} \doteq -0.2$,
because observations contain contributions from both the giant
and the nebula. Therefore, for the purpose of our analysis
we adopted the giant's spectral type of K4, which is
characterized with indices $B-V = 1.41$ and $U-B = 1.66$
\citep[][]{johnson66}.

Applying our method (Eqs. (\ref{eqn:te}) to (\ref{eqn:vg})) to
these data we obtained the electron temperature of the nebula,
  $T_{\rm e} \approx 20\,000$\,K
and its emission measure,
  $EM \sim 1.7 \times 10^{59}$\,cm$^{-3}$.

\subsubsection{AG~Peg}

\astrobj{AG Peg} is known as the slowest symbiotic nova. Currently,
it displays all signatures of a classical symbiotic star
in a quiescent phase \citep[][]{k+93,mn94}. We investigated
\astrobj{AG Peg} around its optical maximum in 1993 November, at the
orbital phase $\varphi = 0.63$. Selection of these photometric
data was important to compare our results with another analysis
and to use appropriate corrections for emission lines
\citep[see][]{lines}. 
The $U$, $B$, $V$ magnitudes were taken from \citet[][]{tomtom98}.
We used values of 6 measurements obtained between
JD~2\,449\,974.3 and JD~2\,450\,024.3 and then averaged our
results obtained from individual nights. For the spectral type
of the giant in \astrobj{AG Peg} we adopted M3 \citep[][]{m+s99}.

Analyzing these data we revealed
   $T_{e}\sim 17\,910$\,K
and
   $EM\sim 4.85\times 10^{59}$\,cm$^{-3}$,
which are quite similar to those obtained by \citet[][]{sk05}.

\subsubsection{Z~And}

\astrobj{Z And} is a prototype of the class of symbiotic stars. We
analysed $U$, $B$, $V$ measurements taken around a light
maximum of a quiescent phase (i.e. $\varphi \sim 0.5$),
because the corrections for emission lines were made at
the same orbital position of the binary. For the purpose
of our analysis, we used $U$, $B$, $V$ magnitudes taken by
\citet[][]{bel92}. We selected 4 measurements from
JD $\sim$ 2\,445\,171.51 to JD $\sim$ 2\,445\,204.42.
The spectral type of the giant in \astrobj{Z And} is M4.5
\citep[][]{m+s99}. However, the giant's colour indices
are available only for spectral types of M4 and/or M5.
Therefore, we used indices for a M4 and M5 giant separately,
and then averaged our results obtained from individual nights.

Our resulting temperature, $T_{e}\sim 29\,300$\,K, is somewhat
higher than that determined by \citet[][]{sk05}. Using indices
of a M5 giant led to a lower temperature,
  $T_{e}\sim 28\,390$\,K,
while indices for a M4 giant yielded
  $T_{e}\sim 30\,220$\,K.
Our emission measure, $EM\sim 7.8\times 10^{59}$\,cm$^{-3}$,
is quite similar to that determined independently
\citep[][]{sk05}.

\begin{table*}
\caption[]{Physical parameters of the nebula and the giant for
           selected symbiotic stars and novae obtained by our
		   method of disentangling their $UBV$ magnitudes.}
\begin{center}
{\footnotesize
\begin{tabular}{ccccccccccc}
\hline
\hline
Object      & Julian date & $T_{e}$ & $T_{e,min}$ - $T_{e.max}$ & $k_{n}$ & $k_{n,min}$ - $k_{n,max}$ & $EM$ & $EM_{min}$ - $EM_{max}$ & $V^{giant}$ & $V_{min}^{giant}$ - $V_{max}^{giant}$ & ph.$^{\star}$\\
            & $JD-2\,4...$& [K] & [K] & [$10^{15}$\,cm$^{-5}$] & [$10^{15}$\,cm$^{-5}$] & [$10^{59}$\,cm$^{-3}$] & [$10^{59}$\,cm$^{-3}$] & [mag] & [mag] &\\
\hline
AX~Per      & 49\,600  & 30\,770 & 27\,600 - 37\,600 & 0.41 & 0.39 - 0.47 & 1.47 & 1.40 - 1.68 & 11.15 & 11.08 - 11.20 & A,E \\ 
\hline
AG~Dra      & 52\,765  & 20\,710 & 12\,330 - 35\,410 & 1.24 & 0.70 - 1.90 & 1.80 & 1.02 - 2.75 & 9.59  & 9.53 - 9.66 & Q \\
            & 53\,172  & 19\,230 & 11\,200 - 32\,610 & 1.16 & 0.62 - 1.79 & 1.68 & 0.90 - 2.59 & 9.57  & 9.50 - 9.64 & Q \\
            & 52\,919  & 44\,940 & 34\,210 - 64\,160 & 13.2 & 11.2 - 16.1 & 19.1 & 16.2 - 23.3 & 9.49  & 9.39 - 9.61 & A \\
\hline
AG~Peg      & 49\,305  & 17\,910 & 16\,500 - 19\,130 & 6.33 & 6.00 - 6.93 & 4.85 & 4.60 - 5.30 & 8.56  & 8.51 - 8.60 & Q \\
\hline
 Z~And      & 45\,293  & 29\,300 & 24\,370 - 38\,040 & 2.89 & 2.59 - 3.14 & 7.78 & 6.99 - 8.45 & 10.08 & 10.01 - 10.16 & Q \\
\hline
RR Tel      & 51\,836  & 18\,910 & 16\,370 - 21\,860 & 0.81 & 0.73 - 0.88 & 6.02 & 5.44 - 6.60 & 11.83 & 11.74 - 11.90 & Q \\
\hline
V1016 Cyg   & 46\,045  & 24\,480 & 22\,100 - 26\,530 & 2.59 & 2.52 - 2.71 & 26.6 & 25.8 - 27.9 &   -   & - & Q \\
\hline
V1974 Cyg   & 48\,883  & 35\,760 & 31\,480 - 41\,420 & 20.4 & 19.3 - 21.8 & 76.5 & 72.3 - 81.8 &   -   & - & Q \\
\hline
\end{tabular}
}
\end{center}
{\footnotesize $^{\star}$~phase: active phase (A), eclipse (E), quiescent phase (Q)}
\label{table:results}
\end{table*}
\begin{figure*}
\begin{center}
\begin{tabular}{cc}
\resizebox{7.7cm}{!}{\includegraphics[angle=270]{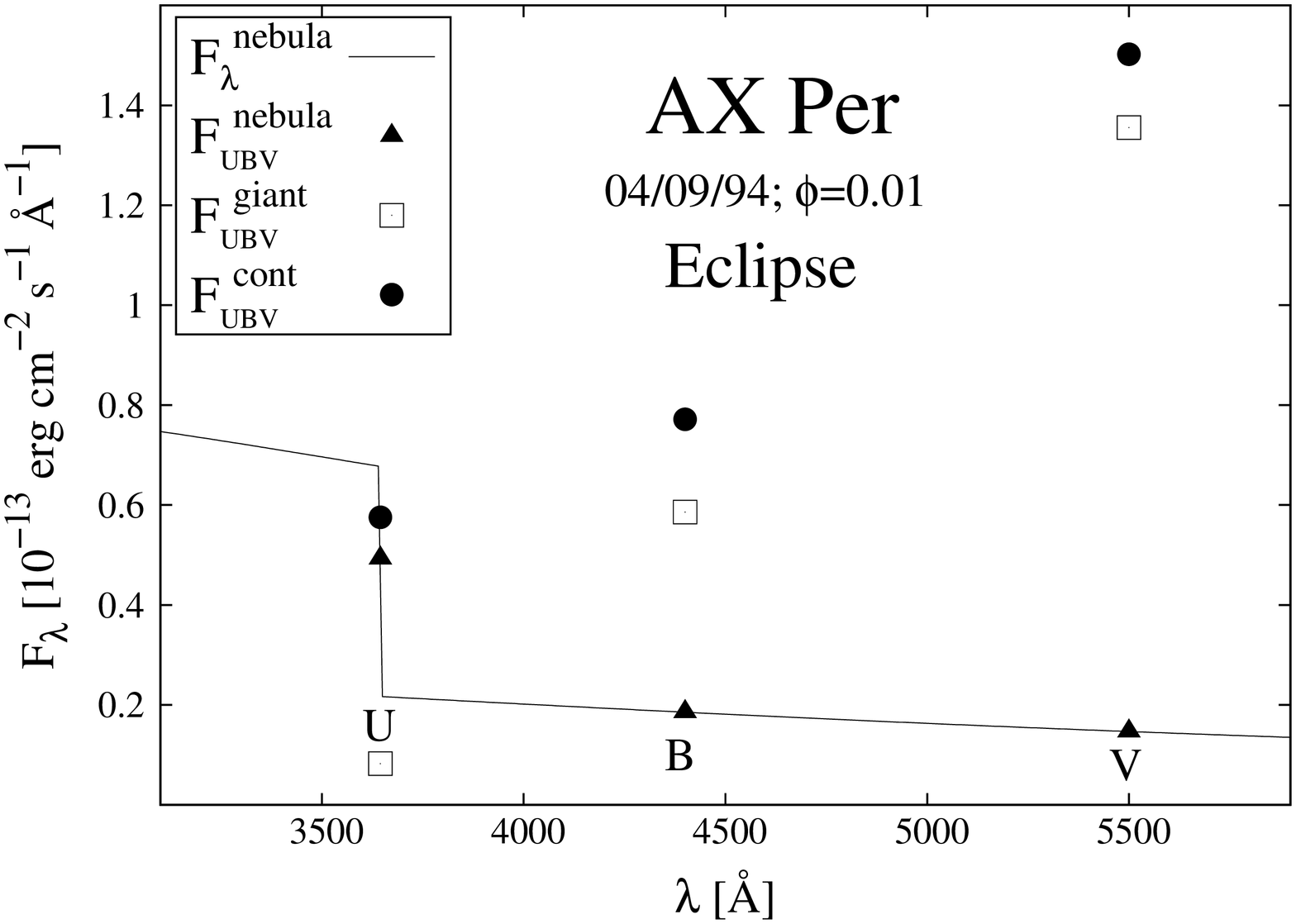}}&
\resizebox{7.7cm}{!}{\includegraphics[angle=270]{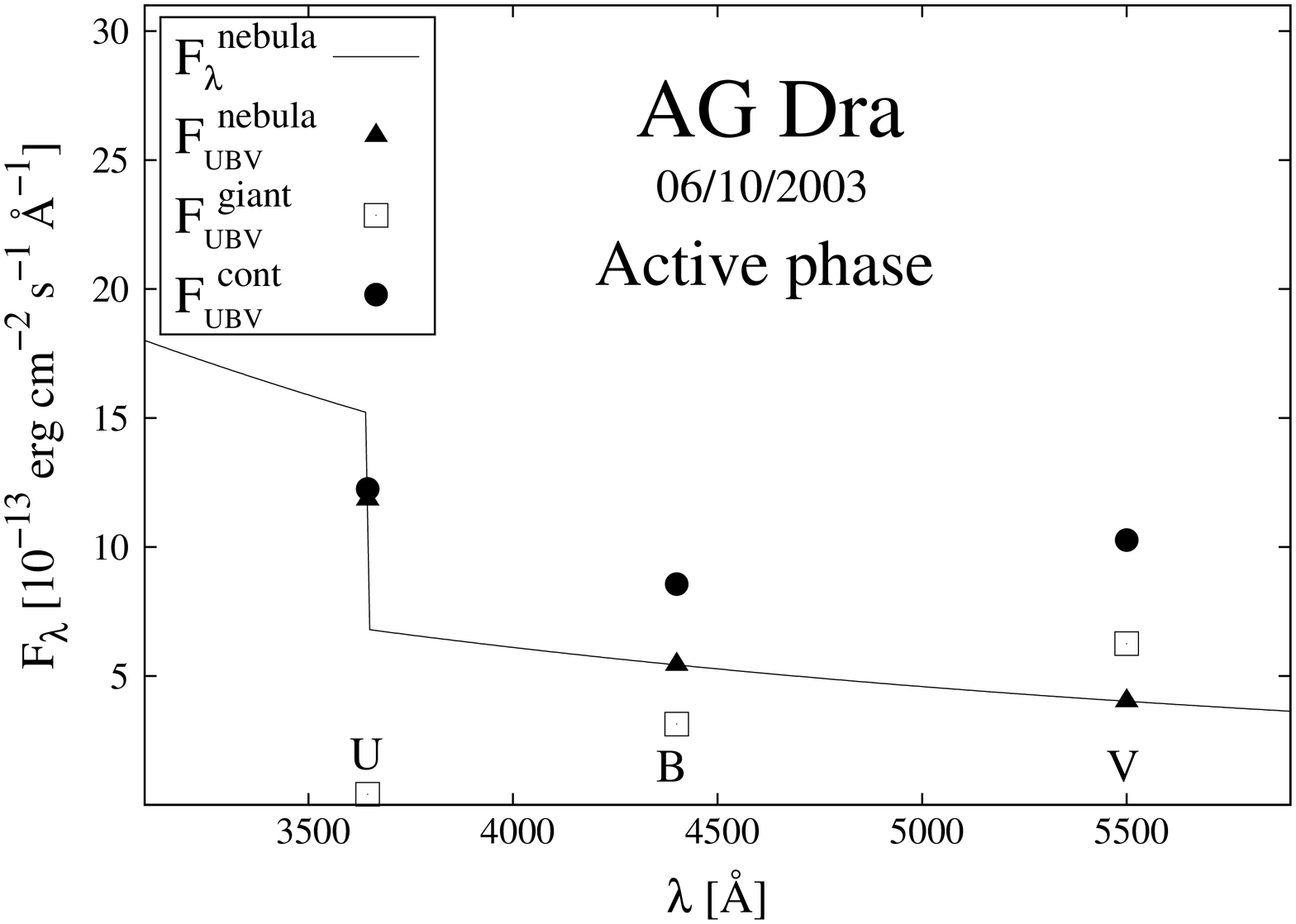}}\\
\resizebox{7.7cm}{!}{\includegraphics[angle=270]{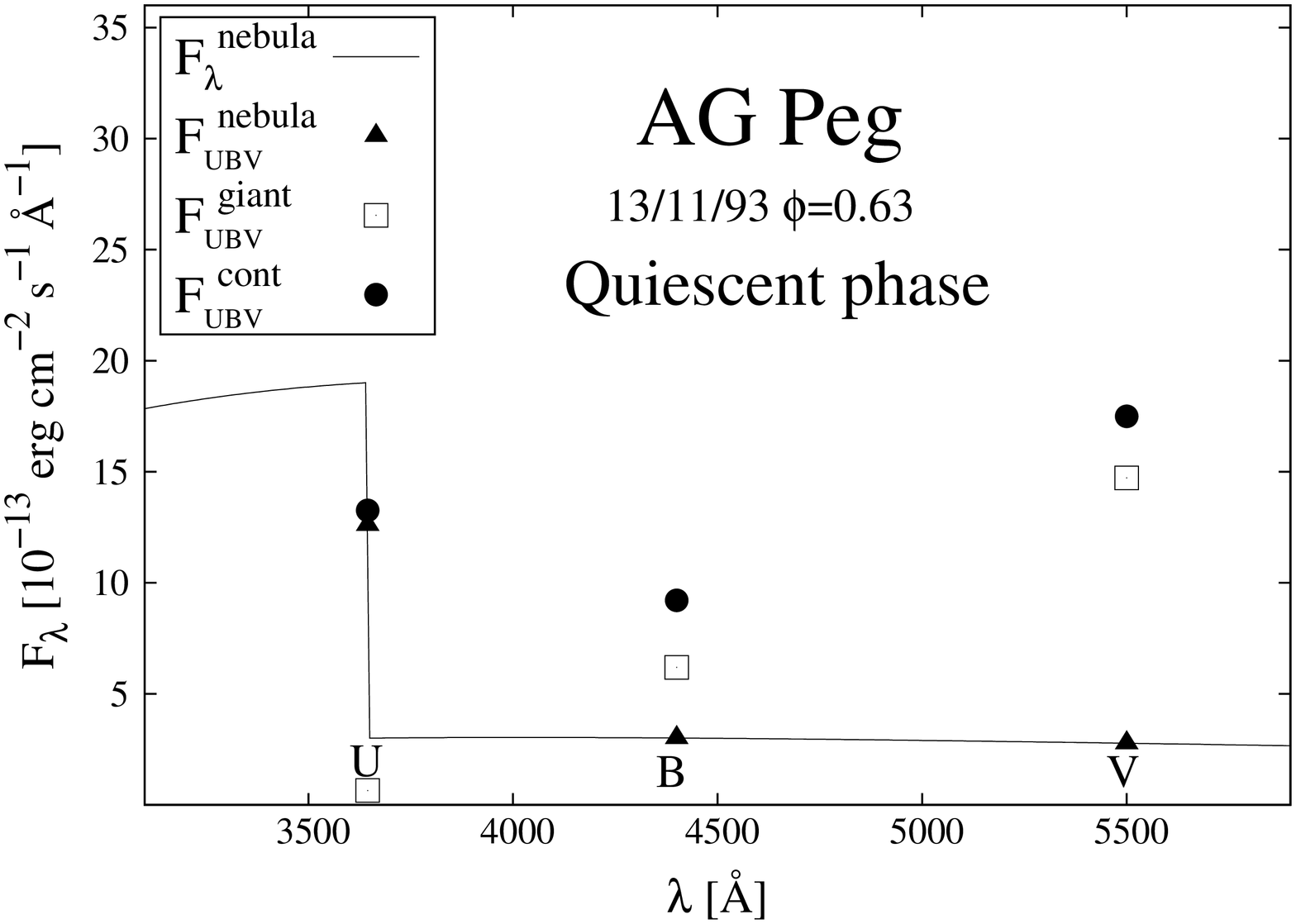}}&
\resizebox{7.7cm}{!}{\includegraphics[angle=270]{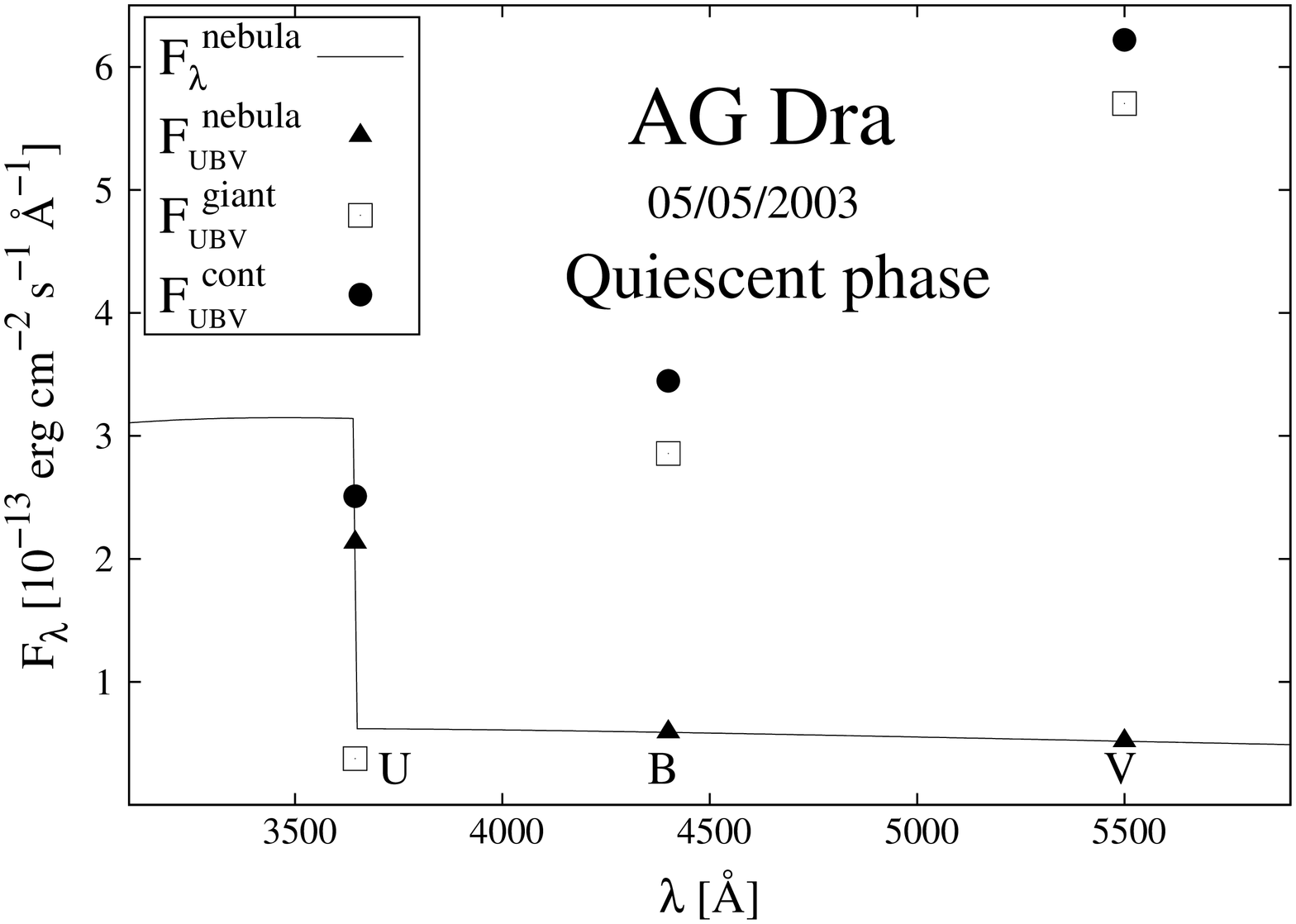}}\\
\resizebox{7.7cm}{!}{\includegraphics[angle=270]{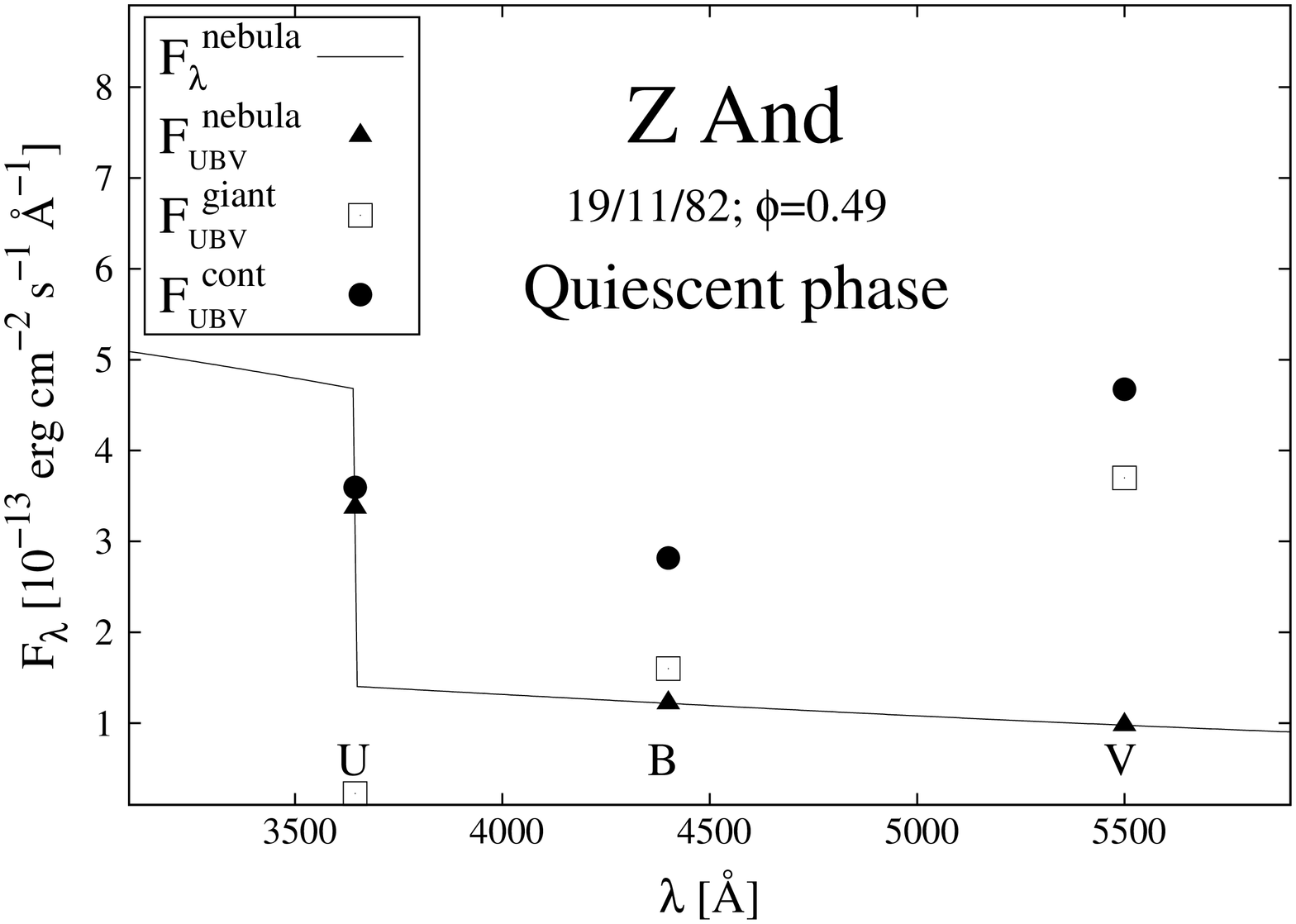}}&
\resizebox{7.7cm}{!}{\includegraphics[angle=270]{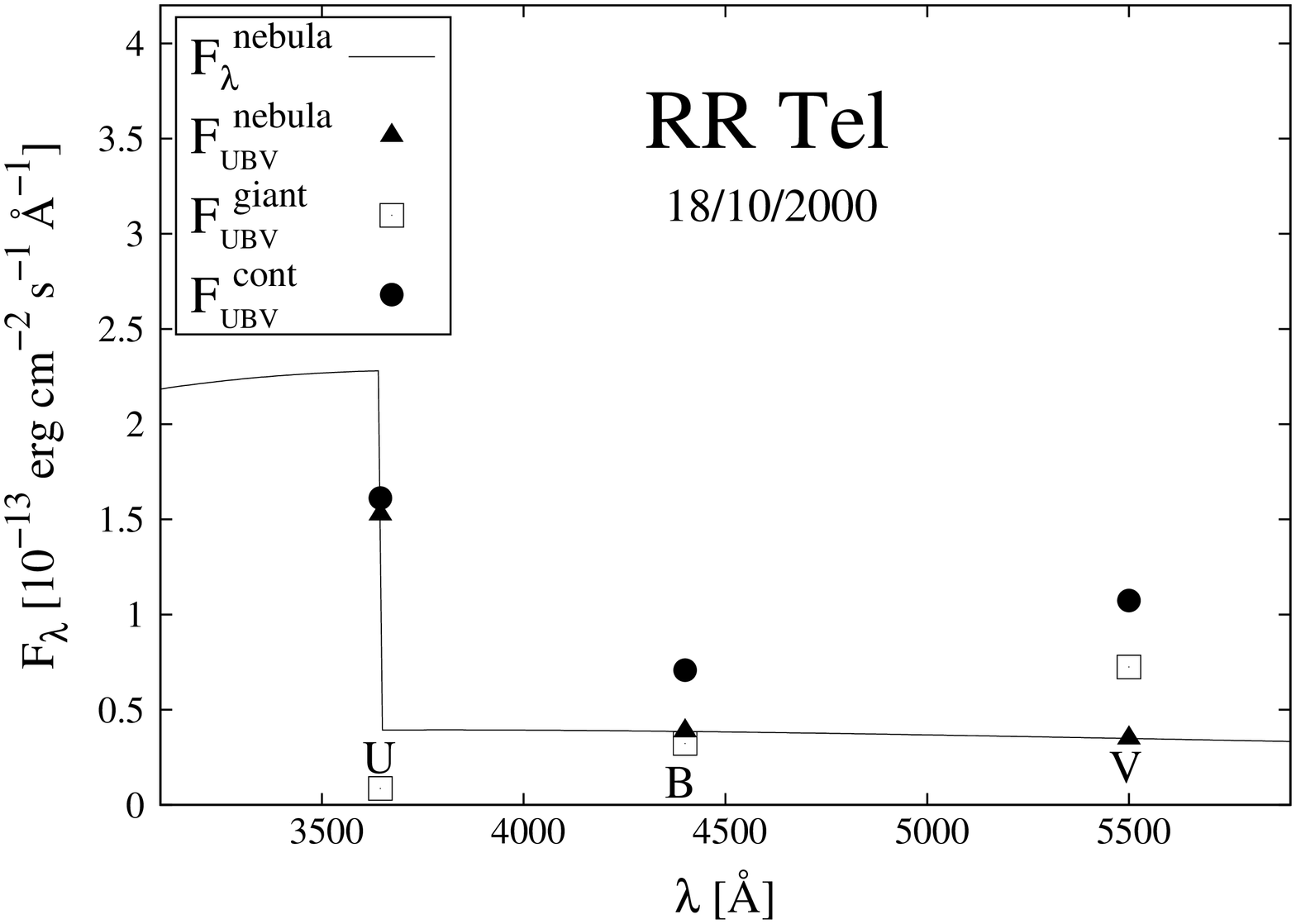}}\\
\resizebox{7.7cm}{!}{\includegraphics[angle=270]{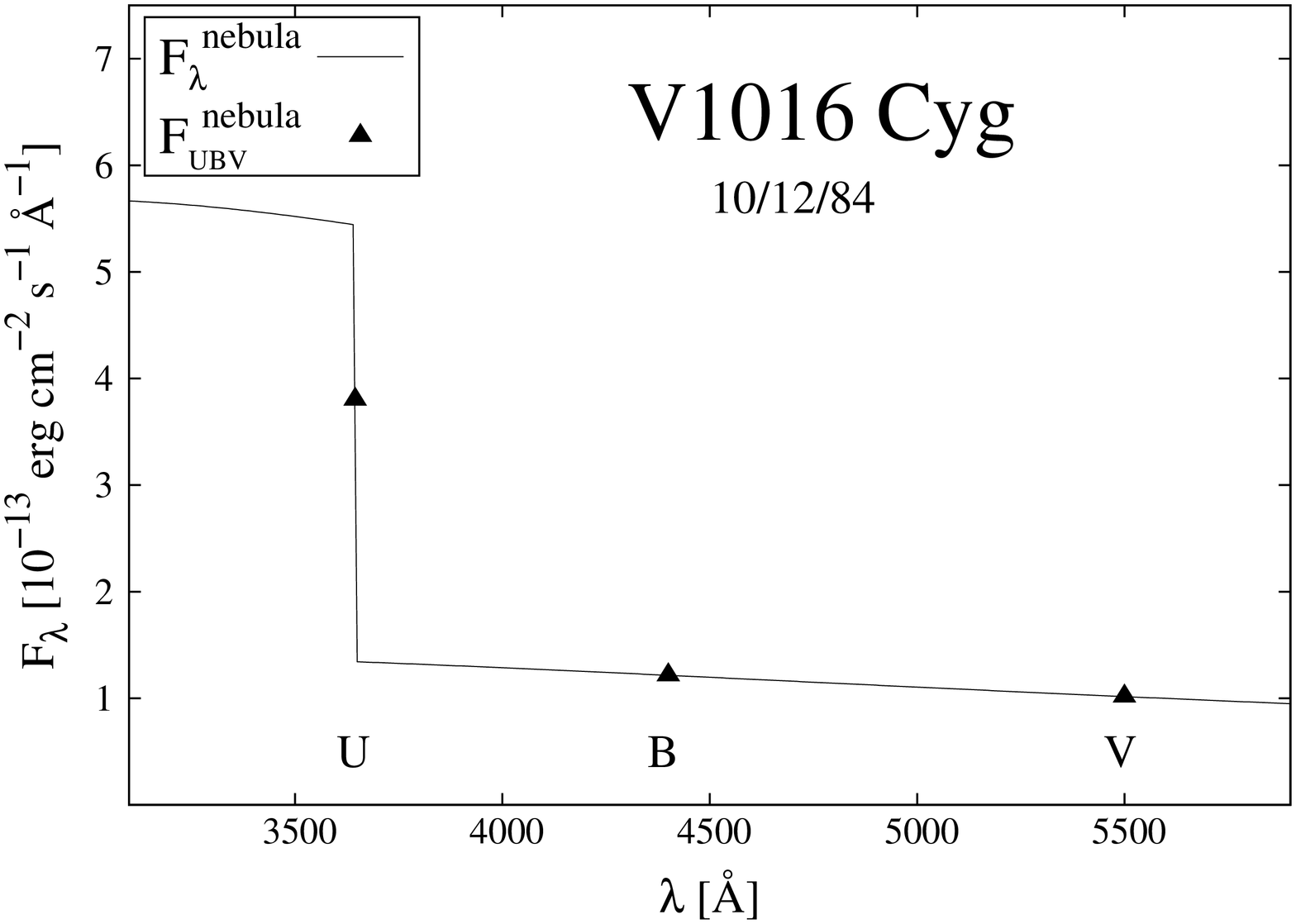}}&
\resizebox{7.7cm}{!}{\includegraphics[angle=270]{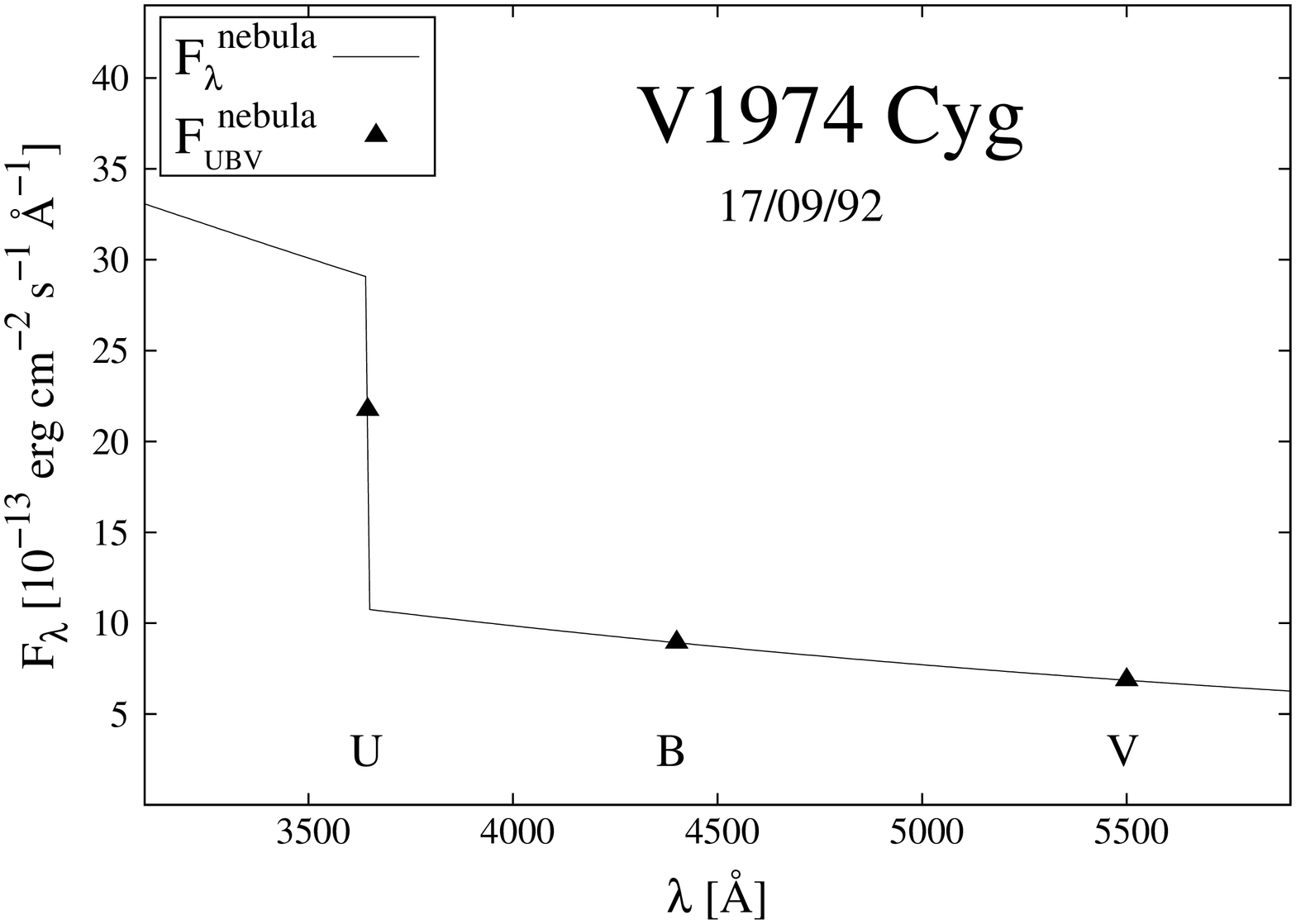}}\\
\end{tabular}
\end{center}
\caption[]{The $UBV$ SED for investigated symbiotic binaries and novae.
The $UBV$ fluxes in the continuum, $F_{\rm UBV}^{cont}$ (Sect.~2.2), and their
disentangled components from nebula, $F_{\rm UBV}^{nebula}$, and the giant,
$F_{\rm UBV}^{giant}$, are denoted by symbols in the legend. The nebular
continuum, $F_{\rm \lambda}^{nebula}$, is drawn with a solid line.}
\end{figure*}

\subsection{Classical symbiotic stars during active phase}

\subsubsection{AX~Per}

\astrobj{AX Per} is known as eclipsing symbiotic binary with
an orbital period of 680 days \citep[][]{sk91}.
The cool component of the binary is a normal giant of the
spectral type M4.5 \citep[][]{m+s99}. During active phases
we observe narrow minima in the light curve at the position
of the inferior conjunction of the giant. It is believed
that they are caused by eclipses of the hot object by
the cool giant.

In this paper we studied the 1994 eclipse, observed during
the 1989-95 active phase of \astrobj{AX Per}. We used 6 $UBV$
measurements from the totality, between JD~2\,449\,570.54
and JD~2\,449\,620.56 \citep[][]{photosymb6}. We assumed
that contributions from both the nebula and the giant had
not been varying significantly during the relatively short
time of the total eclipse ($\sim$ 2 months).
Therefore, also in this case, we derived parameters for each
night set of $UBV$ magnitudes, and then used their means,
to get the most probable values of the fitting parameters.
As in the case of \astrobj{Z And} we applied our method to both
spectral types of a giant, M4 and M5, and used their means
to obtain resulting parameters.

In this way we received the electron temperature of the nebula
  $T_{\rm e}\sim 30\,770$\,K,
the scaling factor,
  $k_{\rm n}\sim 0.41\times 10^{15}$\,cm$^{-5}$
which corresponds to the emission measure,
  $EM\sim 1.47\times 10^{59}$\,cm$^{-3}$.
Using the giant of the spectral type M5 led to lower
values of both nebular parameters than for a M4 giant.
\citet[][]{sk05} modelled the UV--IR SED from the 1990 eclipse
and found very similar value of $T_{\rm e}$, but a factor of
$\sim$2 higher $EM$ than we obtained for the 1994 eclipse.
The lower amount of the nebular emission indicated during
the 1994 eclipse is probably connected with the ending 
of the 1989-95 active phase. 
Note that the high temperature nebula appears to be strong 
just during active phases of symbiotic stars \citep[][]{sk05}. 
Finally, the mean value of the giant's $V$ magnitude is 11.15\,mag. 
However, its value runs from 11.02 to 11.30\,mag for these 
6 individual $UBV$ measurements, including their uncertainties. 
Different values of the $V^{giant}$ magnitudes are beyond
the measured uncertainties. Therefore, we conclude that
they can be caused by a variable brightness of the giant
during the eclipse, perhaps due to its pulsations.

\subsubsection{AG~Dra}

Here we demonstrate our method on the $UBV$ magnitudes measured
during the optical burst of \astrobj{AG Dra} in 2003 October
\citep[see Fig.~8 in][]{sk+07}. We derived
  $T_{\rm e} \approx 45\,000$\,K
and
  $EM \sim 1.9 \times 10^{60}$\,cm$^{-3}$,
which are significantly higher than those we indicated
during quiescent phase.
To check sensitivity of these parameters to the contribution
from the giant, we used different colour indices of the giant.
By this way we obtained a lower $T_{\rm e}\sim 30\,000$\,K and
38\,000\,K for a K2 and K3 giant, respectively.
However, the emission measure was rather insensitive to such
a small change in the spectral type of the giant, because
of its large quantity. We note that during active phases
the emission measure of the symbiotic nebula in \astrobj{AG Dra}
increased by a factor of $\sim$10.

In both cases (quiescence and activity) our parameters
were in a good agreement with those
obtained independently by a precise modelling the $(0.12 - 5)\,\mu$m
SED in the continuum \citep[see Fig.~13 of][]{sk05}.

\subsection{Symbiotic novae}

\subsubsection{RR~Tel}

\astrobj{RR Tel} is a slow symbiotic nova that underwent an outburst
in 1944 and whose optical light curve is still fading
\citep[][]{mn94,kotnik+06}. The binary contains a Mira-type
variable as the cool component, which produces a massive wind
and give rise to a strong dust emission
\citep[e.g][]{jurkic,angeloni}.
The system contains a very hot and luminous white dwarf
\citep[][]{mn94}, which ionizes a fraction of the giant's
wind resulting in a strong nebular emission in the spectrum
of \astrobj{RR Tel} \citep[][]{b+k91}.

For the spectral type of the giant in \astrobj{RR Tel} we adopted M6
\citep[][]{m+s99}.
We applied our method of disentangling the $UBV$ magnitudes
to the measurements made on 1990/06/23 and 1990/06/24 as
published by \citet[][]{munari92}.
We found the electron temperature and the emission measure
that are typical for the quiescent phases of other symbiotic
stars (Table \ref{table:results}).

\subsubsection{V1016~Cyg}

\astrobj{V1016 Cyg} underwent its nova-like outburst in 1964
\citep[][]{Fitzgerald}. The binary contains a Mira variable
as the cool component with a strong IR dust emission
\citep[e.g.][]{ts00,parimucha2002,archipova08}.
The spectral type of the giant in \astrobj{V1016 Cyg} is M7
\citep[][]{m+s99}. However, \citet[][]{johnson66} doesn't
provide the indices for this spectral type. Therefore, firstly
we used the spectral type M6, and secondly, we neglected the
giant contribution to $UBV$ magnitudes, at all. We found
that there is no significant difference between these two
possibilities. As a giant of the spectral type M7 is even
cooler than that of the M6 type (i.e. its contribution to
the optical is lower), we decided to neglect the giant
contribution and calculated only the nebular continuum.
Also, it is important to note that the nebular contribution
dominates the optical \citep[see Fig.~5 of][]{lines}, which
makes it difficult to extract the relatively very faint
light from the giant by our method.
In our analysis we used $UBV$ measurements published
by \citet[][]{parimucha2000} from 1984 October, which
are close to dates of the spectroscopic observations
used for emission lines corrections.
%
The large emission measure, $EM\sim 10^{60}$\,cm$^{-3}$, is 
typical for the active symbiotic systems.

\subsection{Classical nova V1974~Cyg}

\astrobj{V1974 Cyg} (Nova Cygni 1992) belongs to the group of the classical
novae. It was discovered by \citet{collins} on 1992, February 19.
To investigate the physical conditions in this nova by our method,
we used the data from its nebular phase taken around day 210 after
the optical maximum.
According to the X-ray light curve of \astrobj{V1974 Cyg}
\citep[see Fig.~1 of][]{krautter}, the burning white dwarf
in the nova was already very hot at that time. Therefore
we could neglect its contribution to the optical as well as
that from the red dwarf companion.
The optical spectrum was thus strongly dominated by the
nebular radiation. As a result, we derived parameters
characterizing the nebular radiation component only. For this
purpose, we used two flux-points corresponding, for
example, to $U$ and $B$ magnitudes. Particularly, we used
photometric measurements taken at JD~2\,448\,883.4
(1992/09/17) as published by \citet[][]{chochol93}.

Disentangling these $U$ and $B$ measurements, we found
a relatively high electron temperature of
  $T_{\rm e}\sim 36\,000$\,K,
and also a very high emission measure,
  $EM\sim 8\times 10^{60}$\,cm$^{-3}$ (Table~4). 

\section{Discussion}

Physical parameters for the symbiotic nebulae, which we derived by 
disentangling the $UBV$ measurements (Sect.~3, Table~4), are 
in a good agreement with those determined by a precise modelling
the UV--IR SED \citep [][]{sk05}. We found that the emission
measure can be determined with a higher accuracy than the electron
temperature of the nebula. The former is given by scaling 
the emission coefficient, and thus can be estimated
within the uncertainty of the photometric measurements,
while the latter is given by the profile of the volume emission
coefficient. In addition, for lower values of electron 
temperatures, to say $T_{\rm e} < 25\,000$\,K, the profile is 
relatively steeper than for higher values of $T_{\rm e}$. 
As a result a lower $T_{\rm e}$ can be determined with a better
accuracy than a higher one, and vice versa.
Taking into account difference in $T_{\rm e}$ between
quiescent and active phases, we can, in general, conclude
that during the quiescence, $T_{\rm e}$ can be determined
more precisely than in the activity, in spite of equal
uncertainties of $UBV$ measurements.

Parameters of the nebula in our model also depend on colour
indices of the cool giant, which values, however, are often
different in different articles. For the purpose of this work
we adopted indices from \citet[][]{johnson66}. Further, it is
important to know the correct spectral type of the giant.
Also here different authors recommend slightly different
spectral types for the giants in the investigated systems.
Furthermore, our analysis revealed that the brightness of
the cool giants can vary within a few times 0.1 mag
(e.g. \astrobj{AX Per} during the 1994 eclipse).
This type of variability can be ascribed to pulsations of
late-type giants, because they represent their typical behaviour.
All these uncertainties cannot be included in our analysis.
Therefore, our resulting magnitudes of the giant can be
considered only as estimates of their mean values.

It is of importance to note that we modelled just the optical 
continuum. So it is inevitable to know corrections for the
influence of emission lines on the $UBV$ magnitudes. 
Uncertainty (ignorance) of these corrections can represent
a significant source of errors of the fitting parameters.

Knowing actual colour indices of the giant and corrections
for emission lines, our system of equations allowed us to
study the effect of uncertainties of the $UBV$ magnitudes
to the fitting parameters. We could easily found extreme 
values of the parameters, as well as their values based 
on all possible combinations of $UBV$ magnitudes. 

During quiescent phases we found the electron temperature 
$T_{e}\sim 20\,000\,K$, while during active phases our method 
indicated higher values of $T_{e}\sim 30\,000 - 40\,000$\,K. 
During quiescence the emission measure was in the order of 
$\sim 10^{59}$\,cm$^{-3}$, while during activity we identified 
its increase to $\sim 10^{60}$\,cm$^{-3}$. For symbiotic novae 
we obtained even higher values of 
$\sim 10^{60} - 10^{61}$\,cm$^{-3}$. 

Finally, we note that our method of disentangling the $UBV$
magnitudes can be applied to any spectrum composed from
a nebular and stellar component of radiation.

\section{Conclusion}

In this contribution we employed a method of disentangling
the composite spectrum of the optical continuum on the basis
of simple multicolour photometric measurements.
Our method (Sect.~2) allowed us to determine the physical
parameters of the main contributing components of the
radiation into the optical region -- the nebula and the
giant. Main results may be summarized as follows.
\begin{enumerate}[(i)]
\item
 On the basis of the $UBV$ photometric measurements, we
 determined the true optical continuum of the selected
 symbiotic (-like) binaries.
\item
 We compared the observed continuum by a model, which includes 
 contributions from the nebula and the giant (Sect.~3). 
 In this way we determined the electron temperature and
 emission measure of the nebula, and the $V$ magnitude
 of the giant.
\item
 Our model parameters are well comparable with
 those determined independently by another method. In particular,
 by a precise modelling the UV--IR SED as introduced by
 \citet[][]{sk05}.
\item
 Our approach thus provides a good estimate of the physical
 parameters of contributing sources of radiation into
 the optical on the basis of a simple $UBV$ photometry.
\end{enumerate}

\section*{Acknowledgment}
This research was supported by a grant of the Slovak Academy of
Sciences, VEGA No. 2/0038/10.

\appendix
\section {Relation for determining the electron temperature}
In this appendix we derive relation for determining the electron
temperature (Eq. \ref {eqn:te}) from the following system of
equations
\begin{equation}
 F_{\rm U}^{giant} + F_{\rm U}^{nebula} = 10^{-0.4 (U^{cont} +q_{\rm U})},
\label{eqn:1}
\end{equation}
\begin{equation}
 F_{\rm B}^{giant} + F_{\rm B}^{nebula} = 10^{-0.4 (B^{cont} +q_{\rm B})},
\label{eqn:2}
\end{equation}
\begin{equation}
 F_{\rm V}^{giant} + F_{\rm V}^{nebula} = 10^{-0.4 (V^{cont} +q_{\rm V})},
\label{eqn:3}
\end{equation}
\begin{equation}
 F_{\rm U}^{nebula} = k_{\rm n}~\varepsilon_{\rm U}(T_{\rm e}),
\label{eqn:4}
\end{equation}
\begin{equation}
 F_{\rm B}^{nebula} = k_{\rm n}~\varepsilon_{\rm B}(T_{\rm e}),
\label{eqn:5}
\end{equation}
\begin{equation}
 F_{\rm V}^{nebula} = k_{\rm n}~\varepsilon_{\rm V}(T_{\rm e}),
\label{eqn:6}
\end{equation}
\begin{equation}
 \frac{F_{\rm U}^{giant}}{F_{\rm B}^{giant}} = 
      10^{-0.4(U^{giant}-B^{giant}+q_{\rm U}-q_{\rm B})}=10^{-0.4(UB+q_{\rm U}-q_{\rm B})},
\label{eqn:7}
\end{equation}
\begin{equation}
 \frac{F_{\rm B}^{giant}}{F_{\rm V}^{giant}} = 
      10^{-0.4(B^{giant}-V^{giant}+q_{\rm B}-q_{\rm V})}=10^{-0.4(BV+q_{\rm B}-q_{\rm V})}.
\label{eqn:8}
\end{equation}\\
At the beginning we have 8 equations (\ref {eqn:1} - \ref {eqn:8}).\\
We express the flux from the giant in the $V$ passband $F_{\rm V}^{giant}$
from Eq. (\ref {eqn:3}) as
\begin{equation*}
   \boxed{F_{\rm V}^{giant} = 
         10^{-0.4(V^{cont}+q_{\rm V})}-F_{\rm V}^{nebula}}
\end{equation*}
and substitute it to the remaining equations. Equations (\ref {eqn:1}),
(\ref {eqn:2}), (\ref {eqn:4}), (\ref {eqn:5}), (\ref {eqn:6}),
(\ref {eqn:7}) do not change. The Eq.~(\ref {eqn:8}) turns into
\begin{equation}
 F_{\rm B}^{giant}= 10^{-0.4(BV+V^{cont}+q_{\rm B})}
                   -10^{-0.4(BV+q_{\rm B}-q_{\rm V})}F_{\rm V}^{nebula}.
\label{eqn:9}
\end{equation}
We have 7 equations now (\ref {eqn:1}, \ref {eqn:2},
\ref {eqn:4}, \ref {eqn:5}, \ref {eqn:6}, \ref {eqn:7}, \ref {eqn:9}).\\
Further, we express the flux from the giant in the $B$ passband
$F_{\rm B}^{giant}$ from Eq. (\ref {eqn:2}) as
\begin{equation*}
    \boxed{F_{\rm B}^{giant} = 
          10^{-0.4(B^{cont}+q_{\rm B})}-F_{\rm B}^{nebula}} 
\end{equation*}
and substitute it to remaining equations. Equations (\ref {eqn:1}),
(\ref {eqn:4}), (\ref {eqn:5}), (\ref {eqn:6}) don't change.
The Eq. (\ref {eqn:7}) turns into
\begin{equation}
 F_{\rm U}^{giant}= 10^{-0.4(UB+B^{cont}+q_{\rm U})}
                   -10^{-0.4(UB+q_{\rm U}-q_{\rm B})}F_{\rm B}^{nebula}.
\label{eqn:10}
\end{equation}
The Eq. (\ref {eqn:9}) turns into
\begin{equation}
\begin{split}
&10^{-0.4(B^{cont}+q_{\rm B})}-F_{\rm B}^{nebula}=\\
&=10^{-0.4(BV+V^{cont}+q_{\rm B})}-10^{-0.4(BV+q_{\rm B}
  -q_{\rm V})}F_{\rm V}^{nebula}.
\end{split}
\label{eqn:11}
\end{equation} 
We have 6 equations now (\ref {eqn:1}, \ref {eqn:4},
\ref {eqn:5}, \ref {eqn:6}, \ref {eqn:10}, \ref {eqn:11}).\\
Further, we express the flux from the giant in the $U$ passband
$F_{\rm U}^{giant}$ from Eq. (\ref {eqn:1}) as
\begin{equation*}
\boxed{F_{\rm U}^{giant}=10^{-0.4(U^{cont}+q_{\rm U})}-F_{\rm U}^{nebula}}
\end{equation*}
and substitute it to remaining equations. Equations (\ref {eqn:4}),
(\ref {eqn:5}), (\ref {eqn:6}), (\ref {eqn:11}) don't change.
The Eq. (\ref {eqn:10}) turns into
\begin{equation}
\begin{split}
&10^{-0.4(U^{cont}+q_{\rm U})}-F_{\rm U}^{nebula}=\\
&=10^{-0.4(UB+B^{cont}+q_{\rm U})}-10^{-0.4(UB+q_{\rm U}
  -q_{\rm B})}F_{\rm B}^{nebula}.
\end{split}
\label{eqn:12}
\end{equation}
We have 5 equations now (\ref {eqn:4}, \ref {eqn:5},
\ref {eqn:6}, \ref {eqn:11}, \ref {eqn:12}).\\
Further, we express the scaling factor of the nebula $k_{\rm n}$
from Eq. (\ref {eqn:4}) as
\begin{equation*}
   \boxed{k_{\rm n} = 
          \frac{F_{\rm U}^{nebula}}{\varepsilon_{\rm U}(T_{\rm e})}}
\end {equation*}
and substitute it to remaining equations. Equations (\ref {eqn:11}),
(\ref {eqn:12}) don't change.
The Eq. (\ref {eqn:5}) turns into
\begin{equation}
F_{\rm B}^{nebula}=
\frac{\varepsilon_{\rm B}(T_{\rm e})}{\varepsilon_{\rm U}(T_{\rm e})}F_{\rm U}^{nebula}.
\label{eqn:13}
\end{equation}
The Eq. (\ref {eqn:6}) turns into
\begin{equation}
F_{\rm V}^{nebula} = \frac{\varepsilon_{\rm V}(T_{\rm e})}
                     {\varepsilon_{\rm U}(T_{\rm e})}F_{\rm U}^{nebula}.
\label{eqn:14}
\end{equation}
We have 4 equations now (\ref {eqn:11}, \ref {eqn:12},
\ref {eqn:13}, \ref {eqn:14}).\\
Further, we express the flux from the nebula in the $U$ passband
$F_{\rm U}^{nebula}$ from Eq. (\ref {eqn:13}) as
\begin{equation*}
   \boxed{F_{\rm U}^{nebula}=
   \frac{\varepsilon_{\rm U}(T_{\rm e})}
        {\varepsilon_{\rm B}(T_{\rm e})}F_{\rm B}^{nebula}}
\end{equation*}
and substitute it to remaining equations. Equation (\ref {eqn:11})
doesn't change. The Eq. (\ref {eqn:14}) turns into
\begin{equation}
F_{\rm V}^{nebula}=
\frac{\varepsilon_{\rm V}(T_{\rm e})}{\varepsilon_{\rm B}(T_{\rm e})}F_{\rm B}^{nebula}.
\label{eqn:15}
\end{equation}
The Eq. (\ref {eqn:12}) turns into
\begin{equation}
\begin{split}
&10^{-0.4(U^{cont}+q_{U})}-\frac{\varepsilon_{\rm U}(T_{\rm e})}{\varepsilon_{\rm B}(T_{\rm e})}
F_{\rm B}^{nebula}=\\
&=10^{-0.4(UB+B^{cont}+q_{\rm U})}-10^{-0.4(UB+q_{\rm U}-q_{\rm B})}F_{\rm B}^{nebula}.
\end{split}
\label{eqn:16}
\end{equation}
We have 3 equations now (\ref {eqn:11}, \ref {eqn:15}, \ref {eqn:16}).\\
Further, we express the flux from the nebula in the $B$ passband
$F_{\rm B}^{nebula}$ from Eq. (\ref {eqn:15}) as
\begin{equation*}
  \boxed{F_{\rm B}^{nebula}=
  \frac{\varepsilon_{\rm B}(T_{\rm e})}
       {\varepsilon_{\rm V}(T_{\rm e})}F_{\rm V}^{nebula}}
\end{equation*}
and substitute it to remaining equations.
The Eq. (\ref {eqn:11}) turns into
\begin{equation}
\begin{split}
&10^{-0.4(B^{cont}+q_{\rm B})}-\frac{\varepsilon_{\rm B}(T_{\rm e})}
{\varepsilon_{\rm V}(T_{\rm e})}F_{\rm V}^{nebula}=\\
&=10^{-0.4(BV+V^{cont}+q_{\rm B})}-10^{-0.4(BV+q_{\rm B}-q_{\rm V})}F_{\rm V}^{nebula}.
\end{split}
\label{eqn:17}
\end{equation}
The Eq. (\ref {eqn:16}) turns into
\begin{equation}
\begin{split}
&10^{-0.4(U^{cont}+q_{\rm U})}-\frac{\varepsilon_{\rm U}(T_{\rm e})}
{\varepsilon_{\rm V}(T_{\rm e})}F_{\rm V}^{nebula}=\\
&=10^{-0.4(UB+B^{cont}+q_{\rm U})}-10^{-0.4(UB+q_{\rm U}-q_{\rm B})}
\frac{\varepsilon_{\rm B}(T_{\rm e})}{\varepsilon_{\rm V}(T_{\rm e})}F_{\rm V}^{nebula}.
\end{split}
\label{eqn:18}
\end{equation}
We have 2 equations now (\ref {eqn:17}, \ref {eqn:18}).\\
Further, we eliminate the flux from the nebula in the $V$ passband
$F_{\rm V}^{nebula}$ from Eqs. (\ref {eqn:17}) and (\ref {eqn:18}).
We get the following equation:
\begin{equation}
\boxed{
\begin{split}
 F_{\rm V}^{nebula}= \frac{10^{-0.4(BV+V^{cont}+q_{\rm B})}-10^{-0.4(B^{cont}+q_{\rm B})}}
 {10^{-0.4(BV+q_{\rm B}-q_{\rm V})}-\frac{\varepsilon_{\rm B}(T_{\rm e})}{\varepsilon_{\rm V}(T_{\rm e})}}=\\
= \frac{10^{-0.4(UB+B^{cont}+q_{\rm U})}-10^{-0.4(U^{cont}+q_{\rm U})}}{10^{-0.4(UB+q_{\rm U}-q_{\rm B})}
\frac{\varepsilon_{\rm B}(T_{\rm e})}{\varepsilon_{\rm V}(T_{\rm e})}-\frac{\varepsilon_{\rm U}(T_{\rm e})}
{\varepsilon_{\rm V}(T_{\rm e})}}
\end{split}}
\label{eqn:19}
\end{equation}
When we multiply Eq. (\ref {eqn:19}) by its denominators and do
some other simple mathematical operations we get
\begin{equation*}
\begin{split}
 & \frac{\varepsilon_{\rm B}(T_{\rm e})}{\varepsilon_{\rm V}(T_{\rm e})}
    \left[10^{-0.4(UB+BV+V^{cont}+q_{\rm U})}-10^{-0.4(U^{cont}+q_{\rm U})}\right]+\\
 &+\frac{\varepsilon_{\rm U}(T_{\rm e})}{\varepsilon_{\rm V}(T_{\rm e})}
    \left[10^{-0.4(B^{cont}+q_{\rm B})}-10^{-0.4(BV+V^{cont}+q_{\rm B})}\right]+\\
 &+10^{-0.4(BV+U^{cont}+q_{\rm U}+q_{\rm B}-q_{\rm V})}
  -10^{-0.4(UB+BV+B^{cont}+q_{\rm U}+q_{\rm B}-q_{\rm V})}=0
\end{split}
\end{equation*}
For the sake of simplicity we divide previous equation by\\
$10^{-0.4(BV+q_{\rm U}+q_{\rm B})}$.
Finally we get the equation for determining the electron temperature 
in a form 
\begin{equation}
\boxed{
\begin{split}
 &\frac {\varepsilon_{\rm B}(T_{\rm e})}{\varepsilon_{\rm V}(T_{\rm e})}
     \left[10^{-0.4(V^{cont}+UB-q_{\rm B})}-10^{-0.4(U^{cont}-BV-q_{\rm B})}\right]+\\
 &+\frac {\varepsilon_{\rm U}(T_{\rm e})}{\varepsilon_{\rm V}(T_{\rm e})}
     \left[10^{-0.4(B^{cont}-BV-q_{\rm U})}-10^{-0.4(V^{cont}-q_{\rm U})}\right]+\\
 &+\left[10^{-0.4(U^{cont}-q_{\rm V})}-10^{-0.4(B^{cont}+UB-q_{\rm V})}\right]=0, 
\end{split}}
\label{eqn:20}
\end{equation}
which we used in this work (Eq.~(15) in the main text). 
After determining the temperature $T_{\rm e}$ from Eq.~(\ref {eqn:20}), 
we can easily determine other parameters, $F_{\rm V}^{nebula}$,
$F_{\rm B}^{nebula}$, $F_{\rm U}^{nebula}$, $k_{n}$,
$F_{\rm U}^{giant}$, $F_{\rm B}^{giant}$ and $F_{\rm V}^{giant}$. 
For example, we can use equations in the frames.
If we are interested only in three fundamental parameters
($T_{\rm e}$, $k_{\rm n}$ and $V^{giant}$) it is better to
determine the scaling factor of the nebula, $k_{n}$, from
\begin{equation}
k_{\rm n}=\frac {F_{\rm V}^{nebula}}{\varepsilon_{\rm V}(T_{\rm e})}
\end{equation}
as we did in the main text.

\end{document}